\documentstyle[12pt,psfig]{article}

\def\Z{{\cal Z}}

\newcommand{\AmS}{{\protect\the\textfont2
\renewcommand{\thesection}{\Roman{section}}
  A\kern-.1667em\lower.5ex\hbox{M}\kern-.125emS}}

\begin{document}
\rightline {DFTUZ 99/11}
\vskip 2. truecm
\centerline{\bf Three and Two Colours Finite Density QCD at Strong Coupling:}
\centerline{\bf A New Look}
\vskip 2 truecm
\centerline { R. Aloisio$^{a,d}$, V.~Azcoiti$^b$, G. Di Carlo$^c$, 
A. Galante$^{b}$ and A.F. Grillo$^d$}
\vskip 1 truecm
\centerline {\it $^a$ Dipartimento di Fisica dell'Universit\`a 
di L'Aquila,  67100 L'Aquila, (Italy).}
\vskip 0.15 truecm
\centerline {\it $^b$ Departamento de F\'\i sica Te\'orica, Facultad 
de Ciencias, Universidad de Zaragoza,}
\centerline {\it 50009 Zaragoza (Spain).}
\vskip 0.15 truecm
\centerline {\it $^c$ Istituto Nazionale di Fisica Nucleare, 
Laboratori Nazionali di Frascati,}
\centerline {\it P.O.B. 13 -  00044 Frascati, (Italy). }
\vskip 0.15 truecm
\centerline {\it $^d$ Istituto Nazionale di Fisica Nucleare, 
Laboratori Nazionali del Gran Sasso,}
\centerline {\it  67010 Assergi (L'Aquila), (Italy). }
\vskip 3 truecm

\centerline {ABSTRACT}
\vskip 0.5truecm

\noindent
Simulations in finite density, $\beta=0$ lattice QCD by means of the
Monomer-Dimer-Polymer algorithm show a signal of 
first order transition at finite temporal size.
This behaviour agrees with predictions of the mean field approximation,
but 
is difficult to 
reconcile with infinite mass analytical solution.
The MDP simulations are considered in detail and 
severe convergence problems 
are found for the SU(3) gauge group, in a wide region of chemical potential.
Simulations of SU(2) model show discrepancies with MDP results as well.
\vfill\eject

\vskip 0.3truecm

The present scenario of finite density $QCD$ is quite disappointing.
Ideas concerning new phases of $SU(N)$ gauge theories at high density
have been recently proposed and tested using 
phenomenological models \cite{wil}, but a direct (theoretical) 
confirmation is, at present, still lacking.
In the last fifteen years several first principle calculations of 
finite density $QCD$ have been tried. The well-known sign problem
in these simulations has prevented in most cases any success
and, until now, no solution is at sight.

The only exceptions in this scenario are the 
Monomer-Dimer-Polymer (MDP) simulations \cite{mdp}. 
Even if the main limitation of this algorithm is that it is 
effective uniquely in the strong coupling limit, it has been able to 
provide reasonable results in the case of $SU(3)$ gauge
group, in the sense that they are not affected from
early onset and are in qualitative agreement with mean
field calculations.
For these reasons they are considered a non trivial test 
for any newly proposed algorithms in finite density $QCD$.

More recently,
 $\beta=0$ $QCD$ has been exactly solved in the limit of 
infinite mass and 
chemical potential \cite{rigorous}. 
In this limit unexpected results have been
obtained concerning the relevance of the phase as well as the 
role of quenched approximation.

In the next section we are mainly concerned in discussing
the compatibility of MDP results with the infinite mass solution
for $SU(3)$. Indeed we have found difficult to reconcile the 
numerical and analytical predictions. 
This led us to reconsider the MDP algorithm more 
carefully, and we found evidences of severe convergence problems in
a wide region of chemical potential.

Lastly we considered the $SU(2)$ case where simulations are not affected
by the sign problem. 
Results using the Grand Canonical Partition Function formalism turn out 
to be in very good agreement with Hybrid Montecarlo 
calculations \cite{hands} while, once again, MDP results \cite{klae}
are inaccurate in the critical region.

We conclude that, in both cases, the MDP algorithm is unable to reproduce
correctly the physics in the more interesting region of chemical
potential. 

\vskip 0.3truecm
\noindent
%**************************************************
\section {$SU(3)$ at infinite coupling}
%**************************************************
\vskip 0.3truecm

The  analytical results available for SU(3) are essentially the mean 
field predictions 
\cite{mf} and
the exact computation of $QCD$ partition function in the infinite mass
limit \cite{rigorous},\cite{lat98}. 
While the former predicts a first order saturation transition at
a value of $\mu$ smaller than $1/3$ of the baryon mass the 
latter shows a richer scenario. Let us recall the main features
of infinite mass $QCD$ \cite{rigorous}:
\begin{enumerate}
\item[$i)$]
at zero temperature ($L_t\to\infty$) the system undergoes a 
first order saturation
transition. In this case the phase of the Dirac determinant is not
relevant (simulations using the modulus of the determinant are exact
in the thermodynamic limit);
\item[$ii)$]
at non zero temperature the phase is relevant but its contribution
to the free energy density is small (simulations using the modulus
of the determinant are almost exact);
\item[$iii)$]
the quenched approximation is very bad, failing to reproduce not only
quantitatively but also qualitatively the true results. It is a much
worse approximation than modulus $QCD$.
\end{enumerate}

Even if the infinite mass model can not provide physical insights 
for the most interesting cases 
we may still use it as a test for numerical 
algorithms in the strong coupling and large mass regime. 
In a previous paper we used
the (analytically calculable)
canonical partition functions  to check 
the Gran Canonical Partition Function simulations for various lattice
sizes \cite{rigorous}. 
We saw that overlap problems are present and, as expected, 
they are more and more severe for larger lattices.
Here we want to use the infinite mass model as a check 
for the MDP approach \cite{mdp}, up to now the only algorithm that
has been able to handle the sign problem of finite density calculations and
has provided results in very good qualitative agreement with mean field
predictions. 
Even if the applicability of the MDP algorithm is restricted to the
strong coupling regime
the strong first order signal observed for $L_t=4$
is the only evidence we have that the mean field approximation has
been able to catch some relevant aspects of finite density physics.

The MDP results \cite{mdp} are somehow puzzling if considered in the light of
the infinite mass solution. In the original work Karsch and Mutter
saw a strong
first order transition for $L_t=4$ (with spatial extent $L_s=4$ and 8) 
and $0.1 < m < 0.7$ while for $m\to\infty$ \cite{rigorous} the number
density is a smooth function of $\mu$ for any non vanishing lattice
temperature.
If we assume the MDP result for granted we have to conclude that, at finite
$L_t$, the transition disappears at some large bare mass $\bar m$.
This possibility seems very unnatural since it
would imply the existence of a (large) physical scale where 
something dramatic should happen, changing completely the system
behaviour and washing out the transition. 

To solve this puzzle we have tried to use the MDP algorithm \cite{tanks}
directly in the large mass regime in order to have data more
easily comparable with the analytic predictions.

The authors of the MDP code noticed in their original paper
that for small masses the algorithm becomes ineffective.
When we tried to use the code for $m > 1.0$ we saw something
similar: 
even if the acceptance rate in the Metropolis update was
reasonable ($\simeq 10 \%$) in the low density phase,
whatever large the value of $\mu$ the system never moved
into the saturated phase.

The same degradation in performances has been observed 
for increasing  lattice size. This is not surprising for
an algorithm based on a global accept/reject step.
What is more surprising is that the degradation seems to
be related only to the value of $L_t$: simulations
for lattices $8^3\times 4$ are feasible but not for
the $smaller$ volume $4^3\times 8$. We have not been able to
simulate any $L_t>4$ lattice.
This behaviour indicates severe slowing down problems at least 
for some choices of the parameters.

The behaviour of MDP code prevented us from completing
our original program. Any direct comparison of MDP at large masses
(and various $L_t$) with the infinite mass model 
was impossible: a severe quantitative test for the
algorithm could not be performed.

Given the impossibility to obtain MDP results for
values of the parameters different from the ones used in the
original paper,
we have repeated the simulations with these same parameters, $i.e.$
 $m=0.1$ and $V=4^4$.

We used start configurations with zero density as well as
saturated configurations and $O(10^6)$ Montecarlo steps
for each value of $\mu$. The first observation was
a very clear signal of hysteresis in the data:
while runs with  zero density start undergo a strong saturation
transition at $\mu=0.69$ (the published result), runs with saturated 
start jump into the zero density phase for $\mu=0.58$ (fig. 1). 
This result casts some doubts on the determination of the critical
point and may well reconcile the MDP results with mean
field predictions (the mean field prediction for this mass, 
$\mu_c^{MF}=0.61$, lies inside the hysteresis region).

If this behaviour signals a first order transition the width of
the region should shrink with increasing statistics, and one should 
observe a two peak
structure in the probability distribution of the number density,
$i.e.$ observe several flip-flops in the Montecarlo history.

We considered runs of up to $O(10^9)$ accepted configurations
(see  fig. 2) for $\mu$ inside the hysteresys region, observing 
the following typical pattern. Starting from
a zero density configuration $n(\mu)$ remains zero until
$\mu\sim 0.69$. At this point the system typically
goes in the saturated phase.
Once the system is in the saturated phase it never goes back.
The same behaviour has been observed (near $\mu\sim 0.58$) for
the saturated start: once in the zero density region
the system never comes back.
Changing the quark mass we only move the position of the
hysteresis cycle unless we reach too small ($m<0.1$) or too
large ($m>1.0$) masses where simulations can not be performed any
more.

Even if we have physical as well as numerical indications that the
present scenario of MDP results implies a pathological
behaviour of the algorithm, we cannot discard the possibility
of a strong first order transition at small mass and finite $L_t$.
To explain the observed Montecarlo data this transition should be 
characterized, at finite volume ($4^4$), by a huge 
energy barrier between the two equilibrium states thus suppressing
considerably the possibility of transitions among them.
In such a case, a first order signal should  survive at 
smaller lattices too, where an exact simulation is possible.

To overcome the sign problem, the only possibility
consists in measuring observables on an
unbiased ensemble ($i.e.$ the ensemble corresponding to $\mu=0$) 
of random generated configurations. Due to
overlap problems, this technique 
allows to compute observables for any value of $\mu$ only if the
generated ensemble contains a number of configurations of the order
of the exponential of the volume \cite{noi2}, \cite{barb}.
From a practical point of view this limits its applicability to
a very small volume, in our case $V=2^4$. 
Clearly this lattice is definitely not
suitable for real physics but we are simply interested in verifying the
presence of any first order transition signal.

We have simulated a $2^4$ lattice at mass
$m=0.1$ diagonalising the propagator matrix \cite{gibbs} and calculating the
coefficients of the Gran Canonical Partition Function. From
these coefficients we can calculate $\Z$ and its derivatives
for any value of $\mu$. In figs. 3 we plot the logarithm of the
modulus and
the phase of an  averaged coefficient as a function of the
statistics. It is clear that, after
$O(10^5)$ configurations, this coefficient has reached a constant and
almost real value. All the other coefficients converge to a real
and positive quantity with equal or smaller statistics and we
considered $10^6$ configurations sufficient to avoid any
overlap problem in our reconstructed partition function.

We calculated the number density and no evidence of coexistence of
two phases was present (figure 4).
Therefore, the MDP results at $L_t=4$
are not supported by $L_t=2$  simulations
at $m=0.1$. 

We conclude that the MDP results for $SU(3)$ have to be reconsidered.
They are difficult to reconcile with the infinite mass 
solution and, from a numerical point of view, they seem not
self consistent in the most interesting region of $\mu$.
The exact determination of the critical point and the much more 
interesting issue of the order of the transition can not be 
addressed with the MDP technique.

%*******************************************************
\section{$SU(2)$ at infinite coupling}
%*******************************************************

In the previous section we have shown quite convincing evidences
of convergence problems for the MDP algorithm applied to
finite baryon density simulations in $SU(3)$; 
we now address the same issue
in the case of $SU(2)$ as gauge group.

The motivation is twofold: firstly, to investigate 
if the problems of the MDP approach 
present in the $SU(3)$ case are universal and independent of the gauge 
group; moreover, the $SU(2)$ gauge group offers us the
possibility of performing direct simulations using other algorithms,
not only for the smallest possible lattice as in $SU(3)$. In fact
since quarks and antiquarks belong to the same (real)
representation the fermion determinant is real and positive also
for non zero chemical potential, and we can recover
the meaning of Boltzmann weight for the exponential of minus the
action.

In order to have a complete set of results, with $\mu$ varying continuously
in a finite range of values, we used the Gran Canonical Partition 
Function (GCPF) scheme. In the $SU(2)$ case it is possible to test
the (non)occurence of the severe drawbacks observed  in $SU(3)$ 
\cite{rigorous}.
In fact, there exist  results of $SU(2)$ theory in the 
strong coupling limit at 
$\mu \ne 0$, obtained using the HMC (Hybrid Monte Carlo) algorithm
in a $4^3 \times 4$ lattice \cite{hands}; this approach, even if not
convenient from a computer resources point of view (the simulation has
to be repeated for each value of $\mu$ considered), is a good workbench
for our GCPF simulations, being the fermion determinant explicitly 
included in the integration measure. We have therefore tested our
results with those in \cite{hands}.

We have performed simulations in the strong coupling limit at 
lattice volumes $4^3 \times 4$, $6^3 \times 4$ and $8^3 \times 4$ at three 
different values of the quark mass $(m=0.1;0.2;0.4)$ measuring 
the number density and the chiral condensate as functions of the 
chemical potential $\mu$.

In these simulations we have diagonalised, for each quark mass value,
$O(1000)$ gauge configurations
generated randomly (e.g. only with the Haar measure of the group) and 
then reconstructed iteratively the coefficients of the fugacity expansion
of the partition function \cite{gibbs} (GCPF coefficients). 
Rounding effects in the determination
of the coefficients for these relatively large lattices have been 
kept under control
using the same procedur developed for the
SU(3) case \cite{noi2}, \cite{rounding}.
At this point a numerical evaluation of the derivatives of free energy
allows the calculation of the observables we are interested in.

In figure 5 we report the number density and chiral condensate
as obtained in our simulations (continuous line) compared with 
the HMC results of \cite{hands} (diamonds).
From these figures it is evident 
that our simulations reproduce the HMC results quite accurately.
The agreement obtained in $SU(2)$ between the GCPF and HMC schemes 
suggests that sampling problems are not present in this case, at least for
the lattices and operators we used.

As a further check of the goodness of GCPF results we have
computed the pion mass in a $6^3 \times 12$ lattice at the
quark mass values we used in our simulations. 
In fact simplified models predict a phase transition (at least at small
temperature) at chemical potential coinciding, in SU(2),
to half of the mass of the lightest baryon of the theory 
(degenerate with the pion at $\mu=0$).

To extract the critical
value of the chemical potential we have used the following criterium.
The number density appears, with increasing volume, to be 
almost zero up to the
critical point, with a linear rise beyond it and flat at large $\mu$
(saturation).
To identify the critical point we have computed 
$\partial n(\mu)/\partial\mu$ for two volumes and defined $\mu_c$
as the position of the first crossing of the curves.
In the infinite volume limit this definition correctly identifies the 
value of $\mu$ where the linear behaviour starts.
In the table we report our critical chemical potential and
half the pion mass for different values of $m$. 

\begin{center}
\begin{tabular}{|c|c|c|}
\hline
\ $m$ & $\mu_c$  & $\frac{m_{\pi}}{2}$ \\
\hline
\ 0.1 & 0.340(4) & 0.3408(7) \\
\ 0.2 & 0.485(5) & 0.4840(6) \\
\ 0.4 & 0.693(5) & 0.6889(5) \\
\hline
\end{tabular}
\end{center}

We can conclude from these data that our predicted
critical chemical potential equals $m_\pi /2$ and
moves with quark mass in the 
expected way, and this behaviour increases the confidence on our
numerical results.

We now compare the GCPF results with MDP ones.
The published results for $SU(2)$ in the MDP scheme
\cite{klae}
are obtained in simulations of $4^3 \times 4$ and $8^3 \times 4$ lattices;
from these simulations we will compare the number density and chiral 
condensate with ours.

In fig. 6 we report the number density as function of the chemical 
potential computed at $m=0.2$ for $L_t=4$ for the three different lattice 
spatial volumes. 
Superimposed to our data we report the number density obtained with the MDP 
algorithm (from figure 6 of \cite{klae}), at the same quark mass 
in a $8^3 \times 4$ lattice. It is evident a marked difference 
between MDP results and those by our simulations, again limited around the 
critical chemical potential as in the $SU(3)$ case. 
In particular our critical chemical potential
is significantly smaller than the one reported in \cite{klae}. 

The largest part of published MDP results concerns the chiral condensate;
in \cite{klae} there are results for different volumes, thus
allowing a more detailed comparision with our results.
We have computed this observable at the same value of the quark mass 
as in \cite{klae} ($m=0.2$) in three lattices: $4^3 \times 4$, 
$6^3 \times 4$ and $8^3 \times 4$. 
In figure 7 we report our and MDP results (from figure 2 of \cite{klae}).

It is evident that for the smaller lattice 
(i.e. $4^3 \times 4$) the MDP data are in good agreement with ours;
MDP results in $8^3 \times 4$ still agree with ours except at $\mu=0.6$ 
(the critical point derived in $\cite{klae}$). The strong finite volume 
effect noticed by the authors of $\cite{klae}$ seems unlikely on the chiral 
condensate at infinite coupling, at least this far from the chiral limit.

In the case of $SU(3)$ gauge group, as seen before, we have found
severe slowing down for the MDP scheme.
For $SU(2)$ gauge group Klaetle and Mutter, as reported in \cite{klae},
have tested the independence of their results on the initial 
configuration only for the $4^3 \times 4$ lattice. 
In this case the results agree at a good level with ours. 
In our opinion the observed discrepancy has to be ascribed to 
convergence problems
of the MDP algorithm, although they arise at volumes larger than 
in the $SU(3)$ case. 
Once again there are serious doubts on the accuracy that the MDP algorithm 
can achieve near the critical region.

\vskip 0.3truecm
\noindent
\section {Conclusions}
\vskip 0.3truecm

The strong first order signal seen using the MDP code
for $L_t=4$ is difficult to reconcile with $i)$
the absence of a phase
transition at finite temperature and infinite mass and
$ii)$ the (reliable) numerical results on $2^4$ lattice.
To solve this discrepancy we tried to repeat the MDP simulations at larger
masses. This turned out to be impossible due to a dramatic drop in
performances at large ($m>1$) masses. We had the same evidence
trying to change $L_t$ from 4 to larger values.
We repeated the simulations at small mass and $L_t=4$ finding
unexpected huge hysteresis signal but not a direct evidence
of two state coexistence.

The peculiar behaviour of the MDP algorithm seems not confined to
the $SU(3)$ case. Indeed MDP $SU(2)$ simulations 
agree well with HMC and  GCPF results except in the critical
region and the discrepancies are more severe with the system volume.

From these evidences we conclude that the MDP algorithm (for
two and tree colours) is not reliable in the most interesting 
region of $\mu$ where the number density varies rapidly
and no conclusion on the presence of a phase transition
can be achieved using this technique.

More in general, we conclude that even the infinite coupling limit 
of finite density QCD,
in principle easier 
to be studied, is still awaiting an efficient simulation scheme.

\newpage
\noindent
{\bf Acknowledgements}
\vskip 0.3truecm

The authors thank F. Karsch for a critical reading of the manuscript.
A.G. thanks F. Karsch for useful discussions and Istituto Nazionale di 
Fisica Nucleare  for a fellowship at the University of Zaragoza.
This work has been partially supported by CICYT (Proyecto AEN97-1680)
and by a INFN-CICYT collaboration.

\newpage
\begin{figure}[!t]\
\psrotatefirst
\psfig{figure=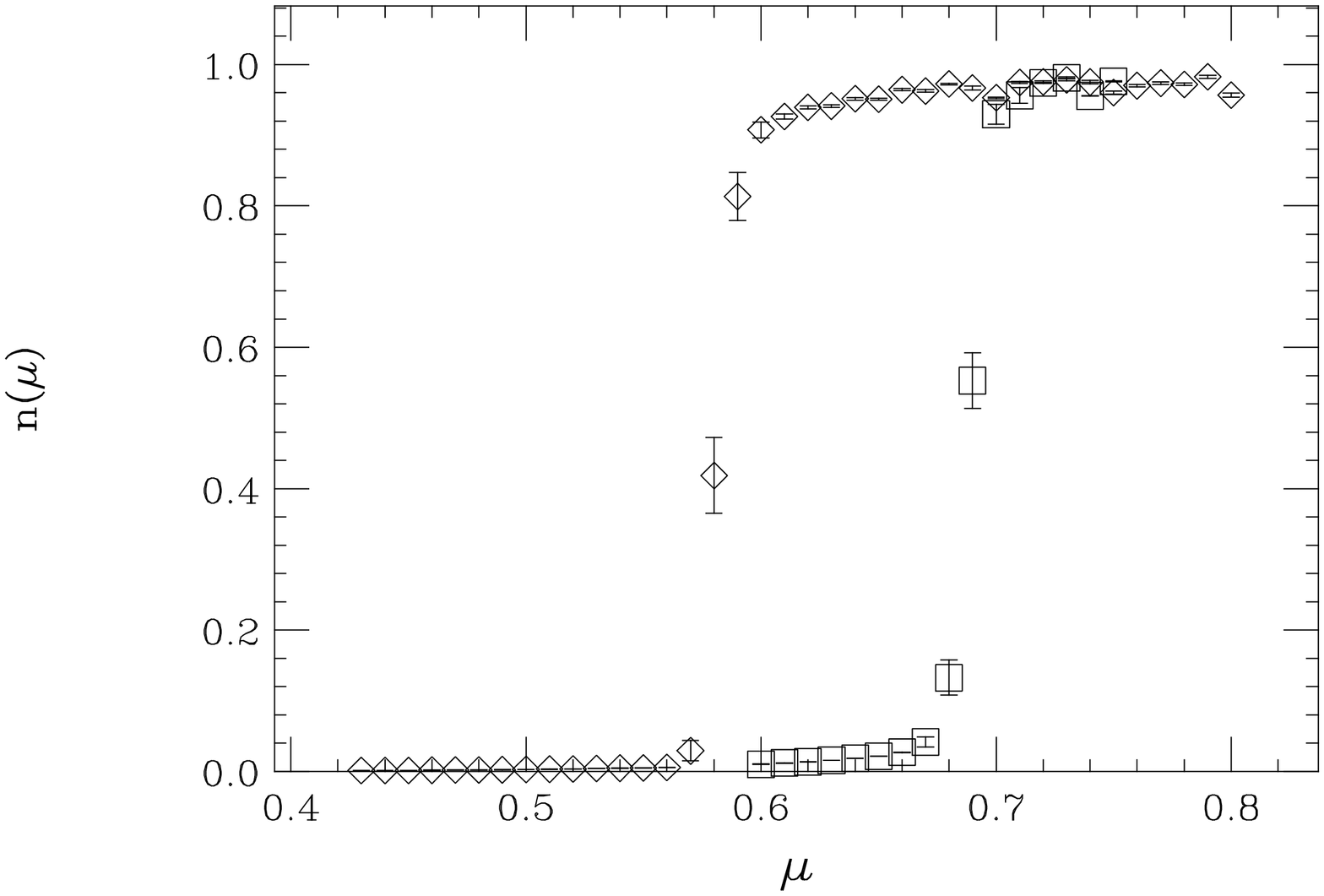,angle=90,width=400pt}
\caption{
Number density for a $4^4$ lattice and $m=0.1$
computed using the MDP algorithm. Zero density starts (squares)
and saturated starts (diamonds).
}
\label{fig1}
\end{figure}

\newpage
\begin{figure}[!t]\
\psrotatefirst
\psfig{figure=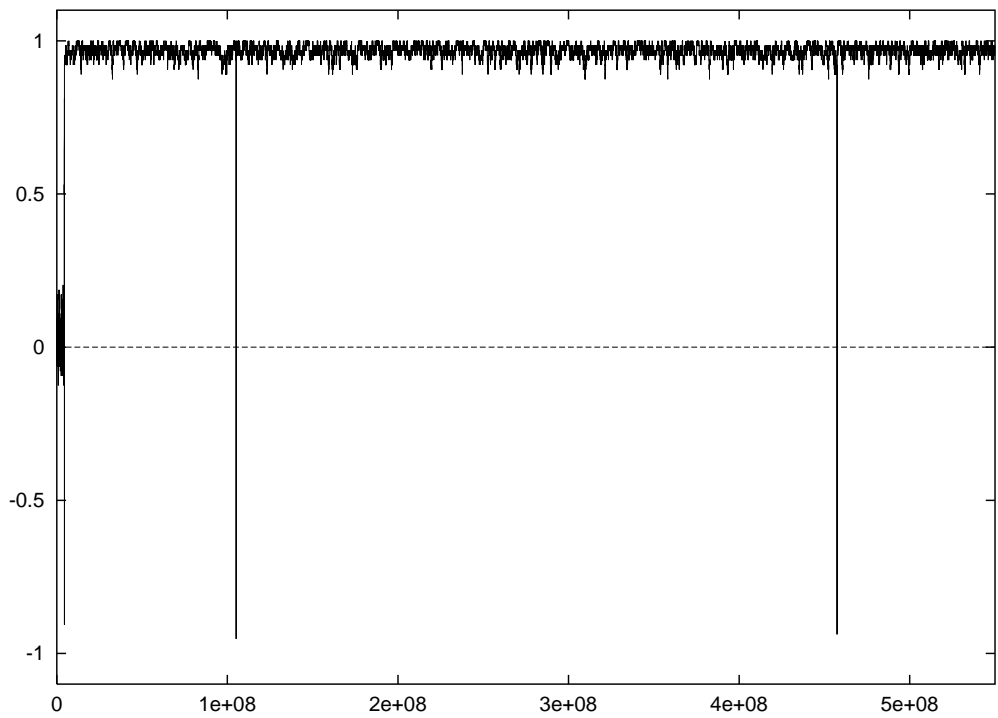,angle=-90,width=400pt}
\caption{
Montecarlo history for the number density in a MDP
simulation at $\mu=0.69$.
}
\label{fig2}
\end{figure}

\newpage
\begin{figure}[!t]\
\hskip 2truecm
\begin{minipage}[t]{75mm}
\psrotatefirst
\psfig{figure=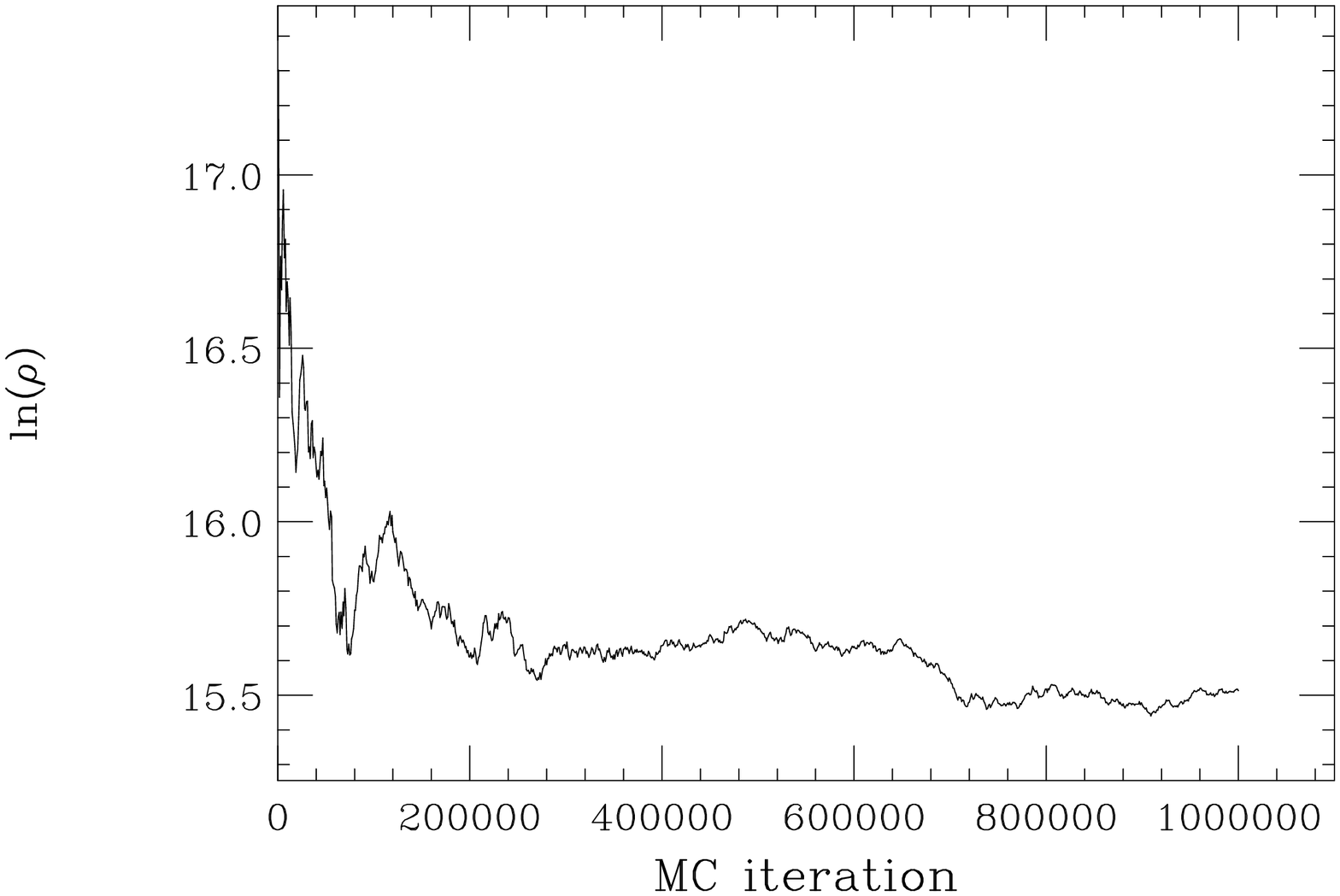,angle=90,width=350pt}
\end{minipage}
\hspace{\fill}
\vskip 0.1truecm
\hskip 2.3truecm
\begin{minipage}[t]{75mm}
\psrotatefirst
\psfig{figure=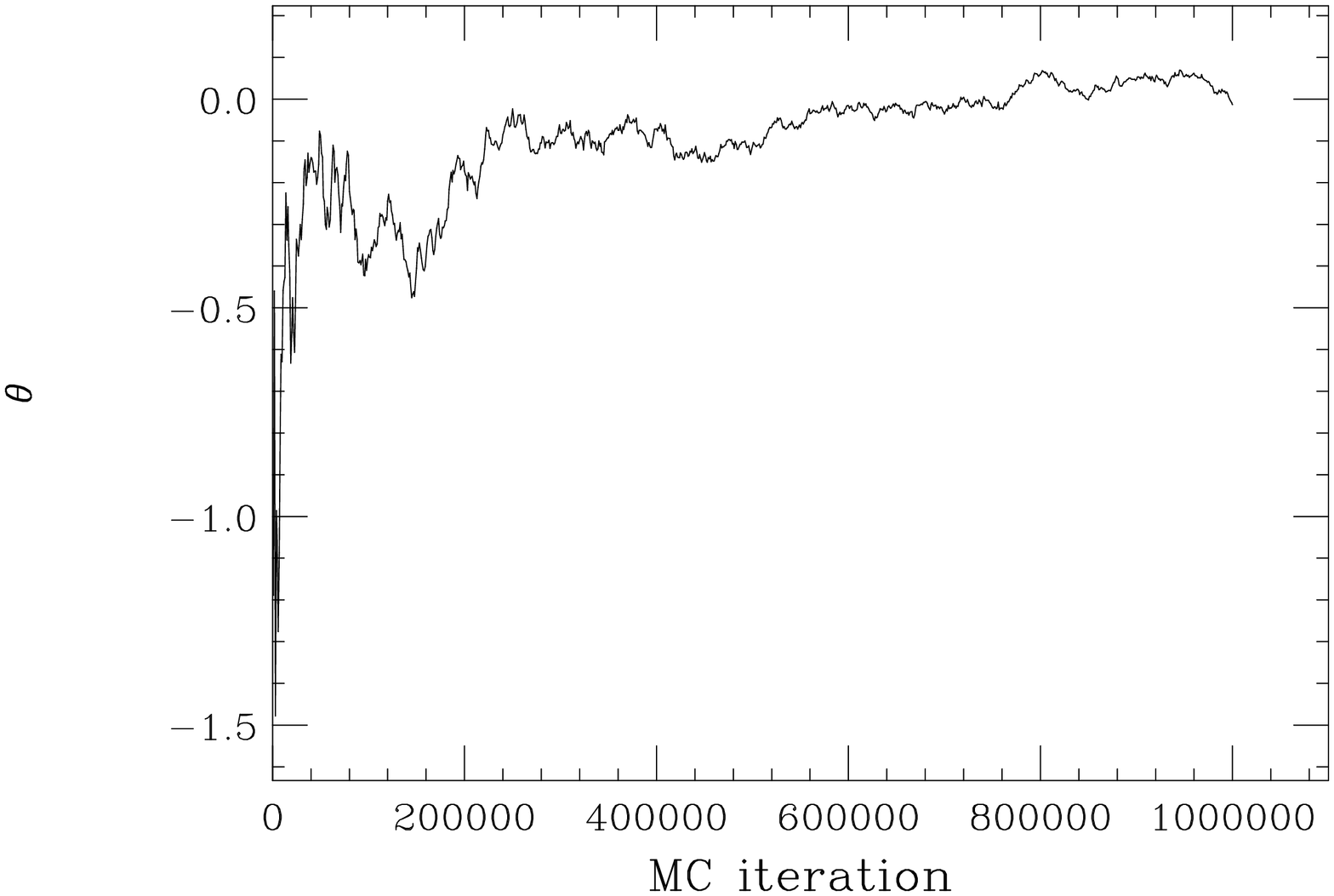,angle=90,width=350pt}
\end{minipage}
\caption{
Plots of the averaged Gran Canonical coefficient of order 72 
$\qquad\quad (<c_{72}>=\rho e^{i\theta})$ as a function of the statistics in a 
$2^4$ lattice at $m=0.1$:
the logarithm of the modulus and the phase.
}
\label{fig3}
\end{figure}

\newpage
\begin{figure}[!t]\
\psrotatefirst
\psfig{figure=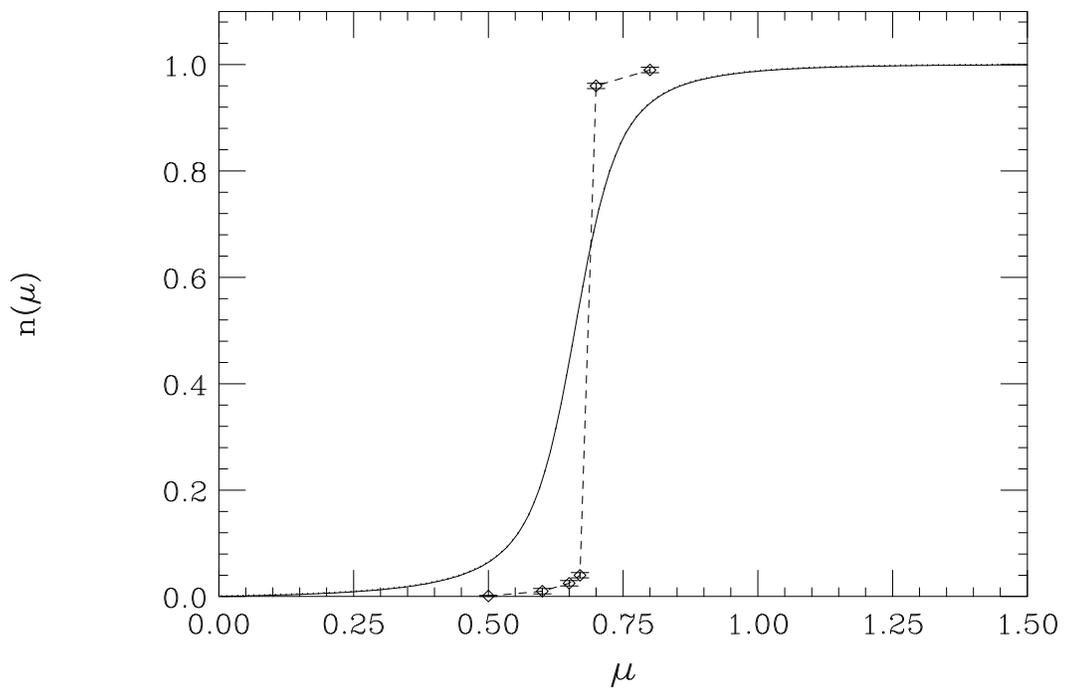,angle=90,width=400pt}
\caption{
Number density at $m=0.1$ evaluated 
using $10^6$ configurations on a $2^4$ lattice
(continuous line) and in a $4^4$ lattice from the
MDP code as reported in [2] (symbols).
}
\label{fig4}
\end{figure}

\newpage
\begin{figure}[!t]\
\hskip 2truecm
\begin{minipage}[t]{75mm}
\psrotatefirst
\psfig{figure=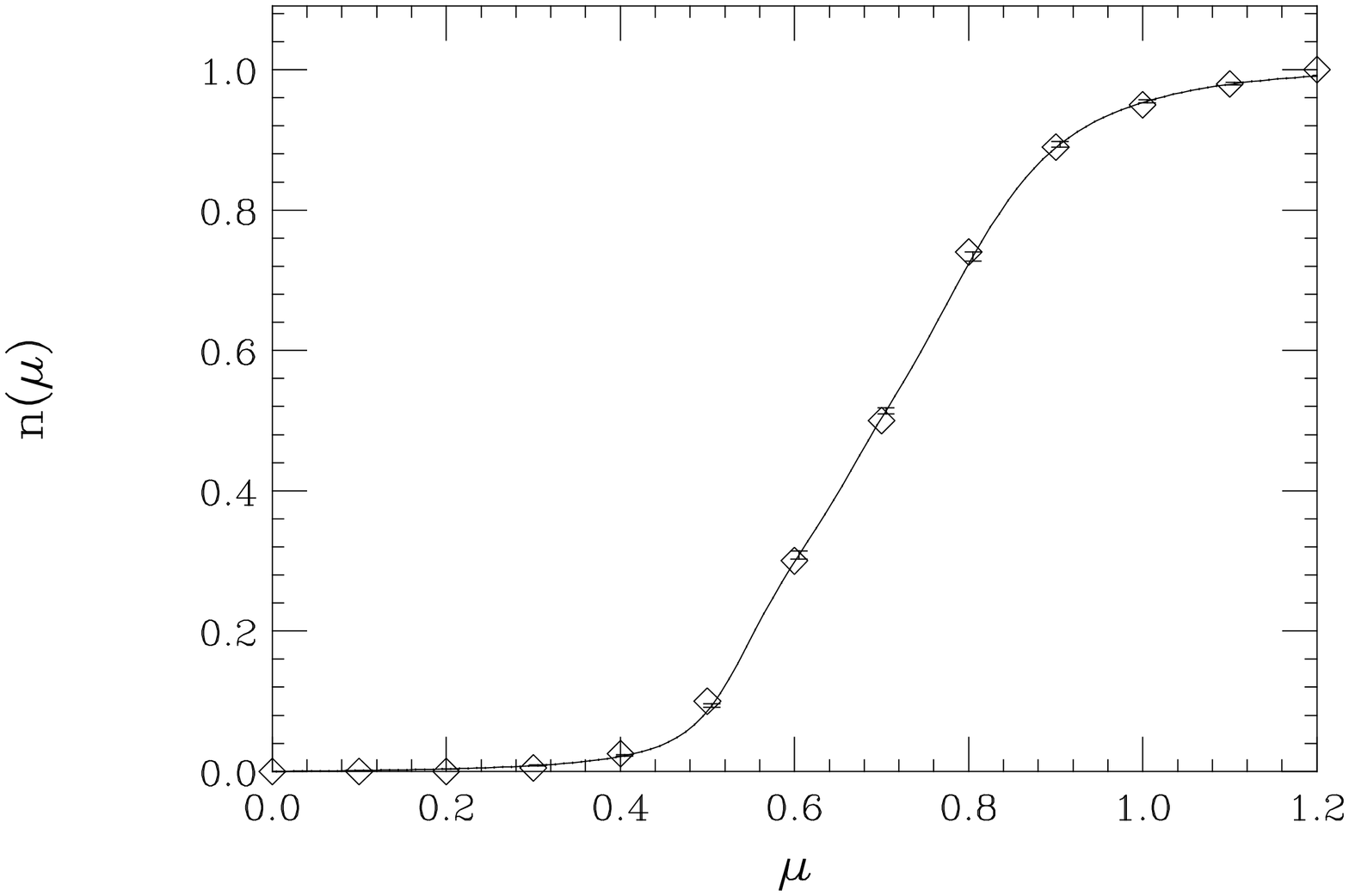,angle=90,width=350pt}
\end{minipage}
\hspace{\fill}
\vskip 0.1truecm
\hskip 2.3truecm
\begin{minipage}[t]{75mm}
\psrotatefirst
\psfig{figure=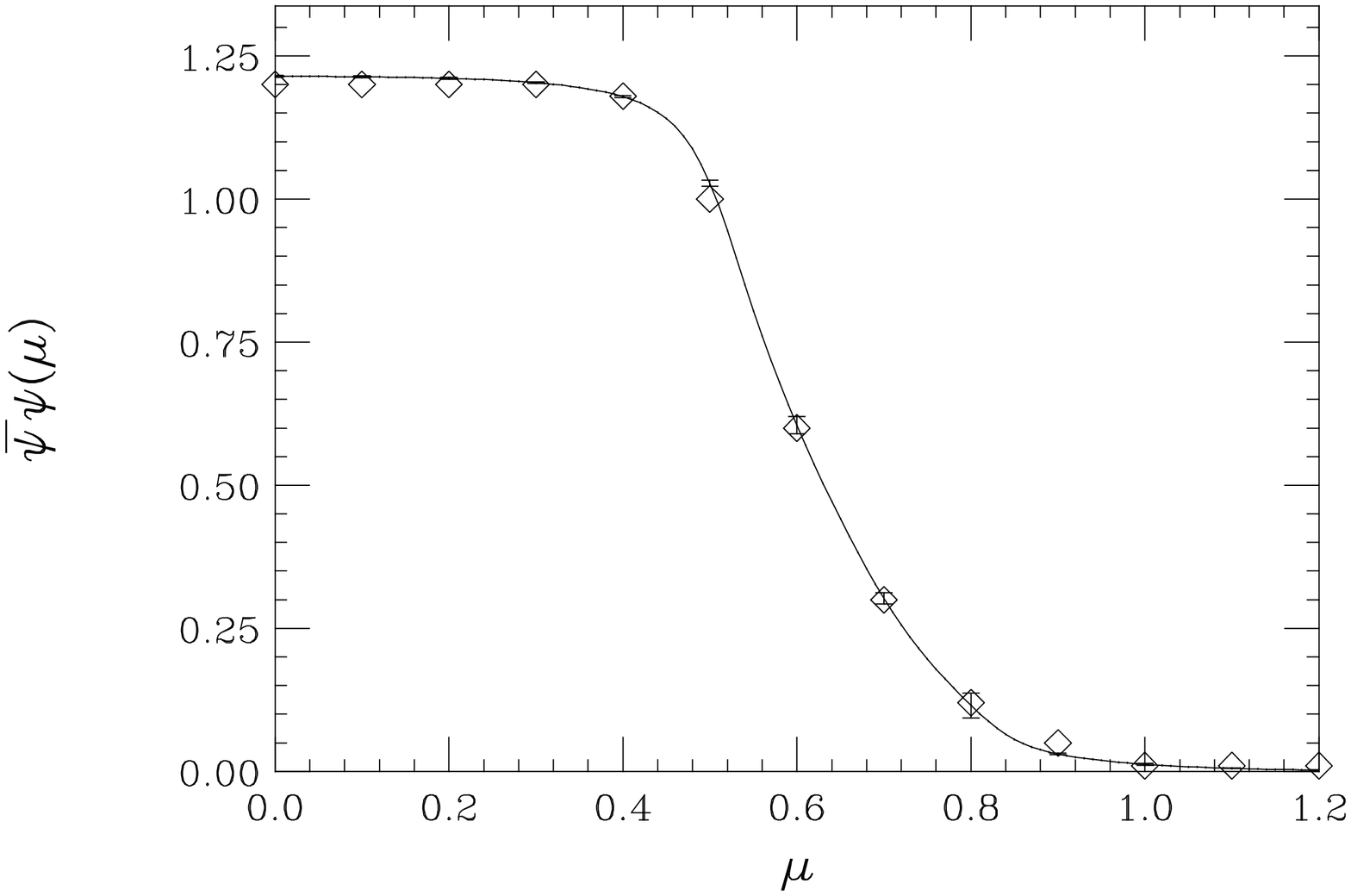,angle=90,width=350pt}
\end{minipage}
\caption{
Number density and chiral condensate at $m=0.2$ in a $4^3 \times 4$
lattice from GCPF (continuous line, the errorbars are reported at some 
values of $\mu$) and from HMC algorithm (diamonds).
}
\label{fig5}
\end{figure}

\newpage
\begin{figure}[!t]\
\psrotatefirst
\psfig{figure=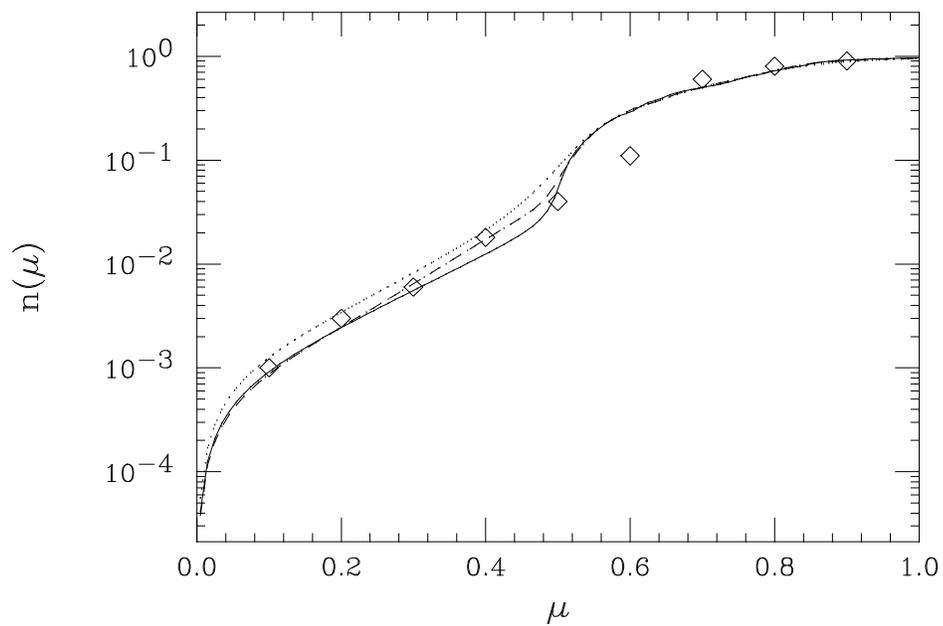,angle=90,width=400pt}
\caption{
Number density computed at $m=0.2$  in the three lattices 
$4^3 \times 4$ (dotted), $6^3 \times 4$ (dashed) and $8^3 \times 4$ 
(continuous)
and from the MDP algorithm (diamonds) in a $8^3 \times 4$.
}
\label{fig6}
\end{figure}

\newpage
\begin{figure}[!t]\
\psrotatefirst
\psfig{figure=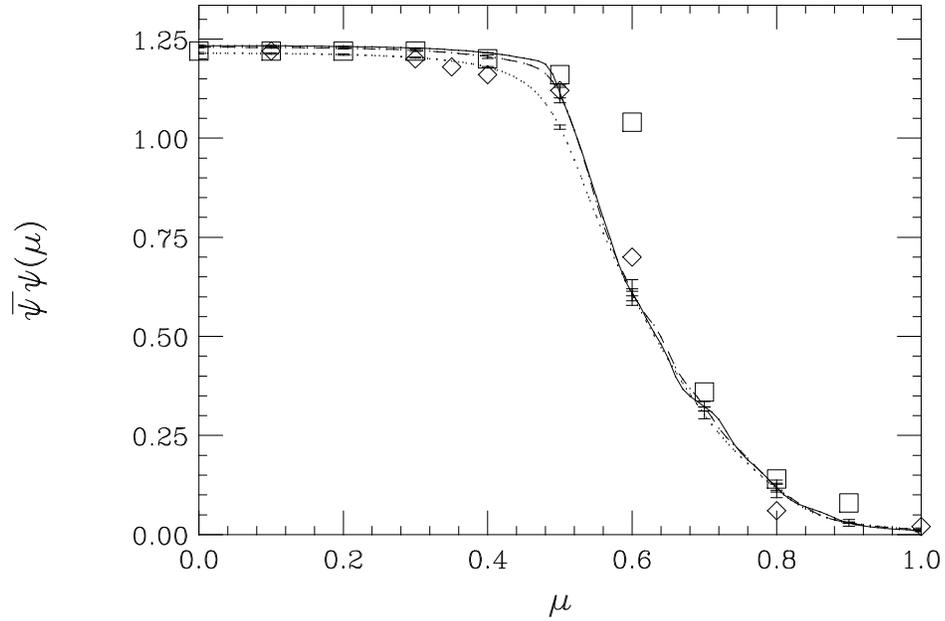,angle=90,width=400pt}
\caption{
Chiral condensate computed at $m=0.2$ in the three lattices 
$4^3 \times 4$ (dotted), $6^3 \times 4$ (dashed) and $8^3 \times 4$ 
(continuous). Errorbars are reported at some 
values of $\mu$. The same quantity for the MDP algorithm in the 
$4^3 \times 4$ 
lattice (diamonds) and in the $8^3 \times 4$ lattice (squares). 
}
\label{fig7}
\end{figure}

\end{document}